\documentclass[amssymb,nofootinbib,nobibnotes,aps,prd,preprintnumbers]{revtex4}
 \usepackage{here}

\usepackage[dvipdfmx]{graphicx}
\usepackage{amssymb,amsfonts,amsmath,bm,color,multirow,cases,empheq,hyperref}
\usepackage{mathrsfs}

\usepackage{enumerate}
\usepackage{ulem}
\usepackage{afterpage}

\hypersetup{%
 setpagesize=false,
 bookmarksnumbered=true,%
 bookmarksopen=true,%
 colorlinks=true,%
 linkcolor=blue,%
 citecolor=red}

\begin{document}




\title{
Multi-centered Myers-Perry Black Holes in Five Dimensions
}

\author{Shinya Tomizawa}
\email{tomizawa@toyota-ti.ac.jp}

\author{Jun-ichi Sakamoto}
\email{jsakamoto@toyota-ti.ac.jp}

\affiliation{Mathematical Physics Laboratory, Toyota Technological Institute,
Hisakata 2-12-1, Nagoya 468-8511, Japan}
\date{\today}

\author{ Ryotaku Suzuki}
\email{suzuki.ryotaku@nihon-u.ac.jp}
\affiliation{Laboratory of Physics, College of Science and Technology, Nihon University,\\
Narashinodai 7-24-1, Funabashi, Chiba 274-8501, Japan}

\preprint{TTI-MATHPHYS-38}




\begin{abstract} 

We present a new family of multi-centered rotating black hole solutions in 5D vacuum Einstein gravity, providing explicit examples of cohomogeneity-three spacetimes.
It is well known that, in the presence of two commuting Killing vector fields, the theory reduces to 3D gravity coupled to an $SL(3,\mathbb{R})$ nonlinear sigma model with five scalar fields.
We show that the scalar fields of the extremal Myers--Perry solution can be expressed in terms of two harmonic functions on 3D flat  space, and that promoting these functions to include multiple sources yields explicit multi-centered extremal Myers--Perry black holes located at arbitrary positions.
Each center forms a smooth $S^3$ Killing horizon, provided that the rotation parameters satisfy $|j_i|<1/2$.
We further demonstrate that all curvature singularities are hidden behind the horizons and that no closed timelike curves arise on or outside the horizons.
The solutions are asymptotically locally Minkowski in the sense that constant-time hypersurfaces are asymptotically locally Euclidean (ALE).
As a concrete example, we consider a binary configuration, examine its rod structure, and demonstrate the absence of conical singularities between the two black holes, indicating that they are supported by an intermediate bubble region separating them.
\end{abstract}

\date{\today}
\maketitle



\section{Introduction}

Exact solutions describing systems of multiple black holes have long been of
interest in both astrophysics and gravitational theory, since they provide
analytic control over configurations that model black-hole binaries---prime
sources of gravitational waves. Constructing such solutions is, however,
notoriously difficult: generic multi-body dynamics are intrinsically
time-dependent and lack enough symmetry to reduce the field equations to a
tractable form. Nevertheless, by imposing additional structure (such as
stationarity, axisymmetry, or integrability in a reduced sigma-model form),
several notable families of static or stationary multi--black--hole geometries
have been found, together with a variety of multi--black-object solutions in
higher dimensions.
A classical four-dimensional example is the Israel--Khan
solution~\cite{Israel1964}, which represents a static, axisymmetric assemblage
of Schwarzschild black holes aligned along a common symmetry axis. In this
configuration, the individual black holes attract each other, and equilibrium
can be maintained only at the cost of introducing conical defects (``struts'')
between the horizons. A rotating analogue was later obtained by Kramer and
Neugebauer~\cite{Kramer1980}, who derived the double-Kerr solution describing
the mutual interaction of two spinning black holes. Although spin--spin
repulsion partly counteracts the gravitational attraction, conical
singularities persist. Subsequent analyses of the double-Kerr family clarified
that achieving regular, balanced configurations in vacuum is highly
nontrivial~\cite{DietzHoenselaers1985,MankoRuiz2001}. These results illustrate a
general theme: within four-dimensional vacuum gravity, exact static or
stationary multi--black-hole configurations typically require distributional
sources (struts/strings) and are generically singular.

\medskip

The situation changes when electric charge is included. The
Majumdar--Papapetrou solution~\cite{Majumdar:1947eu,Papapetrou} provides an exact
static multi--black-hole geometry in the Einstein--Maxwell theory, where the
gravitational attraction is precisely cancelled by the electrostatic repulsion,
yielding completely regular equilibrium configurations without struts. Israel
and Wilson~\cite{Israel:1972vx}, and independently Perj\'es~\cite{Perjes:1971gv},
extended this to stationary settings, producing rotating solutions in
Einstein--Maxwell theory. It was later shown, however, that these spacetimes
contain naked singularities and therefore do not represent genuine black holes.
This establishes that equilibrium configurations of rotating charged black holes
cannot be achieved within Einstein--Maxwell theory alone.
More recently, Teo and Wan~\cite{Teo:2023wfd} constructed a new family of exact,
fully regular multi-centered spinning black holes in 5D
Kaluza--Klein theory. Upon dimensional reduction, these yield balanced
configurations of arbitrarily many dyonic, rotating black holes in
four-dimensional Einstein--Maxwell--dilaton theory. Each constituent black hole
is described by its own mass, angular momentum, and equal electric and magnetic
charges, as well as its position. Moreover, when all angular momenta are set to
zero, the dilaton vanishes and the solution smoothly reduces to the
Majumdar--Papapetrou solution.
More recently, we constructed an exact solution describing multi-centered rotating black holes in 5D Kaluza--Klein theory~\cite{Tomizawa:2025tvb}; upon dimensional reduction, it yields multi-centered rotating black holes carrying both electric and magnetic charges in four-dimensional Einstein--Maxwell--dilaton theory.
This solution provides a rotating generalization of the Majumdar--Papapetrou family and extends the multi-centered rotating black holes of Teo and Wan~\cite{Teo:2023wfd} to the case of unequal electric and magnetic charges.

\medskip

In five dimensions, the landscape of multi--black-hole and multi--black-object solutions is substantially richer than in four dimensions, owing to  the existence of non-spherical horizon topologies. 
A striking demonstration of the richness of 5D gravity is provided by the discovery of black rings~\cite{Emparan:2001wn,Pomeransky:2006bd}. 
The first example of multi--black--object solutions is the ``black saturn'' solution of Elvang and Figueras~\cite{Elvang:2007rd}, which describes a spherical Myers--Perry black hole~\cite{Myers:1986un} surrounded by a black ring. 
This composite object is asymptotically flat and exhibits the remarkable possibility of ``balanced'' configurations without conical singularities, provided the black ring and the central black hole rotate. 
This black saturn, for the first time, showed that regular multi-component stationary solutions can exist in vacuum 5D gravity without supersymmetry.
Furthermore, Iguchi and Mishima constructed the first exact ``black di-ring'' solution~\cite{Iguchi:2007is}, consisting of two concentric black rings in stationary equilibrium. 
By contrast, it remains difficult to construct regular, asymptotically flat vacuum solutions describing multiple black holes with spherical horizons ($S^3$).
A class of 5D, static vacuum solutions describing multiple spherical black holes was obtained by Tan and Teo~\cite{Tan:2003jz}.
Within the generalized Weyl class, they constructed multi-centered configurations of Schwarzschild black holes in five dimensions.
As in the four-dimensional Israel--Khan solution, these spacetimes inevitably contain conical singularities.
Another example is the double Myers--Perry solution found by Herdeiro~\cite{Herdeiro:2008en}, which may be viewed as a 5D analogue of the four-dimensional double-Kerr spacetime.
It describes a pair of rotating black holes, each with an $S^3$ horizon.
As in the double-Kerr case, conical singularities persist and can be interpreted as struts supporting the system, indicating that balanced multi-centered rotating black holes remain difficult to realize even in 5D vacuum gravity.
These works motivate a closer examination of whether fully regular, asymptotically flat multi--black-hole solutions with rotation can exist in the vacuum case, and in particular whether the Myers--Perry black hole admits genuine multi-centered generalizations.

\medskip

In this paper, we construct a new class of multi-centered rotating black hole solutions in 5D vacuum Einstein gravity.
The solutions describe multi-centered extremal Myers--Perry black holes, with independent spin parameters at each center and with the centers located at arbitrary points in 3D Euclidean space.
Our construction assumes only two commuting Killing vectors: an asymptotically timelike Killing vector and a single rotational Killing vector.
Consequently, the solutions are not, in general, of Weyl class (which requires additional axial symmetries), but instead form a cohomogeneity-three family.
Assuming the existence of two commuting Killing vector fields (one timelike and one rotational), the field equations reduce to 3D Euclidean gravity coupled to an $SL(3,\mathbb{R})$ non-linear sigma model whose scalar sector consists of five fields: three inner products of the Killing vectors and two twist potentials.
Using the Maison formulation~\cite{Maison:1979kx} together with Cl\'ement's construction~
\cite{Clement:1986bt,Clement:1985gm}, we show that the scalar fields of the extremal Myers--Perry black hole admit a representation in terms of two harmonic functions on flat $\mathbb{E}^3$.
By promoting the single-source harmonic functions to multi-source ones, we obtain an explicit class of multi-centered extremal Myers--Perry configurations with centers located at arbitrary points in $\mathbb{E}^3$.
We then analyze the geometry in detail.
Each center forms a smooth $S^3$ Killing horizon provided the spin parameters satisfy $|j_i|<1/2$, and we derive the corresponding near-horizon geometry and horizon-area.
We show regularity in the domain of outer communication: all curvature singularities are hidden behind the horizons and closed timelike curves (CTCs) are rigorously excluded.
The solutions are asymptotically locally flat, being asymptotically flat modulo $\mathbb{Z}_N$ (where $N$ the number of black holes) identification, which leads to lens-space asymptotics.
As a concrete example, we study a binary configuration, determine its rod structure, and identify the appearance of an intermediate bubble region, highlighting geometric features distinctive to 5D gravity.

\medskip

We briefly outline the organization of this paper.
Section~\ref{sec:formalism} reviews the Maison formalism and explains how the
5D Einstein equations with two commuting Killing fields can be
recast as a 3D gravity-coupled $SL(3,\mathbb{R})$ sigma model. 
The essential field equations required for the construction of our solution are also collected 
there, following the formulation of Cl\'ement. 
Section~\ref{sec:solution} shows that the extremal Myers--Perry geometry arises 
from a pair of harmonic functions sourced at a single point, and that promoting 
these functions to include multiple centers naturally yields a family of  multi--Myers--Perry configurations. 
In Section~\ref{sec:anaysis}, we investigate the physical and geometric features 
of these solutions---such as their horizon structure, asymptotics, conditions 
for regularity, and the exclusion of CTCs. 
Section~\ref{sec:binary} presents a concrete example of a double Myers--Perry black hole configuration.
Section~\ref{sec:summary} offers concluding remarks and a summary of our results.

\section{ Non-linear sigma model in 5D Einstein gravity with two commuting Killing vectors}\label{sec:formalism}

To study stationary solutions with a single $U(1)$ symmetry, we assume that the spacetime admits two commuting Killing vector fields: a timelike Killing vector and a spacelike rotational Killing vector (at least asymptotically).
Under this symmetry assumption, the 5D Einstein equations can be dimensionally reduced to 3D Euclidean gravity coupled to five scalar fields~\cite{Maison:1979kx}.
We briefly recall how these scalars are organized  as an $SL(3,\mathbb{R})$-invariant nonlinear sigma model.
In general, the reduced equations remain difficult to solve because the sigma-model fields are coupled to the 3D Euclidean metric.
Cl\'ement~\cite{Clement:1985gm,Clement:1986bt} showed, however, that if the 3D base is taken to be flat, one can construct a distinguished class of solutions governed by two harmonic functions.
While Cl\'ement considered the asymptotically Kaluza--Klein case, we will focus on asymptotically flat solutions, or more generally on solutions whose spatial slices are asymptotically locally Euclidean such as the Eguchi-Hanson space.
In what follows, we summarize the equations needed for our analysis, following Cl\'ement's formulation.

\subsection{5D Einstein equation with two commuting Killing vectors}

Let $\xi_a$ $(a=0,1)$ be two mutually commuting Killing vector fields, so that $[\xi_a,\xi_b]=0$ and ${\cal L}_{\xi_a} g=0$.
Introducing coordinates $x^a$ adapted to $\xi_a$ (i.e.\ $\xi_a=\partial/\partial x^a$), the metric $g$ can be written in the  form
\begin{eqnarray}
ds^2 = \lambda_{ab}(dx^a+\omega^a{}_i dx^i)(dx^b+\omega^b{}_j dx^j)
      +|\tau|^{-1}h_{ij}dx^i dx^j \,, \label{eq:5D metric}
\end{eqnarray}
where the scalar fields $\lambda_{ab}$, $\tau:=-\det(\lambda_{ab})$, the functions $\omega^a{}_i$, and the 3D metric $h_{ij}$ $(i=2,3,4)$ are independent of the Killing coordinates $x^a$.
In the vacuum case, there exist locally twist potentials $V_a$ such that
\begin{eqnarray}
\partial_k V_a
=\tau \sqrt{|h|}\,\lambda_{ab}\,\varepsilon_{kij}\,h^{im}h^{jn}\partial_m \omega^b{}_n\,.
\label{eq:domega}
\end{eqnarray}
With these definitions, the vacuum Einstein equations reduce to the field equations for the five scalar fields $\{\lambda_{ab},V_a\}$,
\begin{eqnarray}
\Delta_h \lambda_{ab}
&=&\lambda^{cd} h^{ij}\frac{\partial \lambda_{ac}}{\partial x^i} \frac{\partial \lambda_{bd}}{\partial x^j}
+\tau^{-1} h^{ij}\frac{\partial V_a}{\partial x^i} \frac{\partial V_b}{\partial x^j},
\label{eq:eom1}\\
\Delta_h V_a
&=& \tau^{-1} h^{ij}\frac{\partial \tau}{\partial x^i} \frac{\partial V_a}{\partial x^j}
+\lambda^{bc} h^{ij}\frac{\partial \lambda_{ab}}{\partial x^i} \frac{\partial V_c}{\partial x^j},
\label{eq:eom2}
\end{eqnarray}
together with the Einstein equations for the 3D metric $h_{ij}$,
\begin{eqnarray}
R^h_{ij}
&=&  \frac{1}{4} \lambda^{ab}\lambda^{cd}
              \frac{\partial \lambda_{ac}}{\partial x^i }  \frac{\partial \lambda_{bd}}{\partial x^j }
   + \frac{1}{4}\tau^{-2}\frac{\partial \tau}{\partial x^i} \frac{\partial \tau}{\partial x^j }
    -\frac{1}{2}\tau^{-1}\lambda^{ab} \frac{\partial V_a}{\partial x^i }\frac{\partial V_b}{\partial x^j }\,,
\label{eq:Rij}
\end{eqnarray}
where $\Delta_h$ is the Laplacian and $R^h_{ij}$ is the Ricci tensor associated with $h_{ij}$.

\subsection{Coset matrix}

Maison~\cite{Maison:1979kx} showed that the action for the scalar fields $\{\lambda_{ab},V_a\}$ can be described as a non-linear sigma model with a global $SL(3,\mathbb{R})$ symmetry.
This symmetry becomes manifest upon introducing  the symmetric $3\times3$ matrix
\begin{eqnarray}
\chi= \left(
  \begin{array}{ccc}
  \displaystyle \lambda_{ab}-\frac{V_ aV_b^T}{\tau} &  \displaystyle \frac{V_a}{\tau}\\
   \displaystyle  \frac{V_b^T}{\tau}                     & \displaystyle  -\frac{1}{\tau}
    \end{array}
 \right) \,, \label{eq:chidef}
\end{eqnarray}
where this is  symmetric, $\chi^T=\chi$, and unimodular, $\det(\chi)= 1$.
In terms of $\chi$, the field equations (\ref{eq:eom1}), (\ref{eq:eom2}) and   \ref{eq:Rij}) can be derived from the action
\begin{eqnarray}
S=\int\left(R^h
 - \frac{1}{4}\,h^{ij}\,{\rm tr}\!\left(\chi^{-1}\partial_i\chi \,\chi^{-1}\partial_j\chi\right)
\right)\sqrt{|h|}\,d^3x \,,
\end{eqnarray}
which is invariant under the global transformation
\begin{eqnarray}
\chi\to \chi'=g\,\chi\,g^T,\qquad h\to h \,,
\label{eq:sl3r}
\end{eqnarray}
with $g\in SL(3,\mathbb{R})$.
Equations (\ref{eq:eom1}), (\ref{eq:eom2}) and ~(\ref{eq:Rij}) take the compact form
\begin{eqnarray}
d\star_h (\chi^{-1} d\chi)=0, \label{eq:eomb}\\
R^h_{ij}=\frac{1}{4}\,{\rm tr}\!\left(\chi^{-1}\partial_i\chi \,\chi^{-1}\partial_j\chi\right).
\label{eq:eom2b}
\end{eqnarray}
Thus, the presence of two commuting Killing vector fields reduces 5D vacuum gravity to 3D gravity coupled to an $SL(3,\mathbb{R})$ nonlinear sigma model.
If both Killing vectors are spacelike, the target space is the Riemannian coset $SL(3,\mathbb{R})/SO(3)$, whereas if one of them is timelike, it is replaced by the Lorentzian coset $SL(3,\mathbb{R})/SO(2,1)$.

\subsection{5D asymptotically flat solutions}

Since Eqs.~(\ref{eq:eomb}) and~(\ref{eq:eom2b}) are coupled, solving them in general is nontrivial.
A considerable simplification occurs, however, if the 3D metric $h_{ij}$ is taken to be flat, namely the Euclidean metric on ${\mathbb E}^3$,
\begin{eqnarray}
h_{ij}dx^idx^j=d{\bm x}\cdot d{\bm x},
\label{eq:h}
\end{eqnarray}
where ${\bm x}=(x,y,z)$ are Cartesian coordinates on ${\mathbb E}^3$.
In this case the field equations reduce to
\begin{eqnarray}
\partial_i\!\left(\chi^{-1}\partial^i\chi\right)=0,
\label{eq:eomc}\\
{\rm tr}\!\left(\chi^{-1}\partial_i\chi\,\chi^{-1}\partial_j\chi\right)=0.
\label{eq:eom2c}
\end{eqnarray}
Following Cl\'ement~\cite{Clement:1986bt,Clement:1985gm}, one can represent asymptotically Kaluza-Klein solutions by a coset matrix of the form
\begin{eqnarray}
\chi=\eta\,e^{fA}e^{gA^2},
\label{eq:chi}
\end{eqnarray}
where $f$ and $g$ are harmonic functions on ${\mathbb E}^3$, and $\eta$ and $A$ are a constant $3\times3$ matrices.
Assuming that $f$ and $g$ vanish at infinity, asymptotic flatness of the corresponding 5D metric requires
\begin{eqnarray}
\eta=
\begin{pmatrix}
-1  &  0& 0 \\
  0&0 & -1   \\
0  & -1 &0
\end{pmatrix}.
\end{eqnarray}
which  is different from that of  asymptotically Kaluza-Klein solutions~~\cite{Clement:1986bt,Clement:1985gm}.
Moreover, if $A$ satisfies
\begin{eqnarray}
A^T=\eta A \eta,\qquad {\rm tr}(A)=0,\qquad {\rm tr}(A^2)=0,
\end{eqnarray}
then $\chi$ is symmetric ($\chi^T=\chi$), unimodular ($\det\chi=1$), and automatically obeys the constraint~(\ref{eq:eom2c}).

\section{From an extremal Myers-Perry black hole to Multi-black holes}\label{sec:solution}

The 5D Myers-Perry solution~\cite{Myers:1986un}  describes a rotating black hole, whose horizon cross section has the topology of a three-sphere $S^3$.
It is an asymptotically flat, stationary and bi-axisymmetric solution of the vacuum Einstein equations in five  dimensions.
In this section, we show that the extremal limit can be expressed in terms of the solution with two harmonic functions, Eq.~(\ref{eq:chi}), each having a single point source. 
By extending these harmonic functions to include multiple point sources, we then construct an exact solution describing multi-centered rotating black holes of the 5D vacuum Einstein equations.

\subsection{Extremal Myers-Perry black hole}
The metric for the 5D Myers-Perry black hole is written as~\cite{Myers:1986un}
\begin{equation}
ds^2=-dt^2+\frac{m}{\Sigma}(dt-a_1\sin^2\bar\theta d\phi_1-a_2\cos^2\bar\theta d\phi_2)^2+
(\bar r^2+a_1^2)\sin^2\bar\theta d\phi_1^2+(\bar r^2+a_2^2)\cos^2\bar\theta d\phi_2^2
+\frac{\Sigma}{\Delta}d\bar r^2+\Sigma d\bar\theta^2,
\end{equation}
with the metric functions
\begin{eqnarray}
\Delta&=&\bar r^2\left(1+\frac{a_1^2}{\bar r^2}\right)\left(1+\frac{a_2^2}{\bar r^2}\right)-m,\\
\Sigma&=&\bar r^2+a_1^2\cos^2\bar\theta+a_2^2\sin^2\bar\theta,
\end{eqnarray}
where $m$ and $a_1,a_2$ are the mass and rotational parameters. 
The angular coordinates $(\bar\theta,\phi_1,\phi_2)$ take values in the ranges
\begin{eqnarray}
0\le \bar\theta\le \frac{\pi}{2},\qquad 0\le \phi_1<2\pi,\qquad 0\le \phi_2<2\pi.
\end{eqnarray}
The spacetime admits three mutually commuting Killing vector fields: a timelike Killing vector $\partial_t$ (at least asymptotically) and two rotational Killing vectors $\partial_{\phi_1}$ and $\partial_{\phi_2}$.
At spatial infinity $\bar r\to\infty$, the metric approaches the 5D Minkowski form,
\begin{eqnarray}
ds^2\simeq -dt^2+d\bar r^2+\bar r^2\Bigl(d\bar\theta^2+\sin^2\bar\theta\, d\phi_1^2+\cos^2\bar\theta\, d\phi_2^2\Bigr).
\end{eqnarray}
Introducing new angular coordinates $(\theta,\psi,\phi)$ via
\begin{eqnarray}
\bar\theta=\frac{\theta}{2},\qquad \phi_1=\frac{\psi+\phi}{2},\qquad \phi_2=\frac{-\psi+\phi}{2},
\end{eqnarray}
we see that $\partial_\psi$ and $\partial_\phi$ are also Killing vectors.
In these coordinates, the $S^3$ metric at spatial infinity takes the standard Hopf fibration form, exhibiting $S^3$ as an $S^1$ bundle over $S^2$:
\begin{eqnarray}
ds^2\simeq -dt^2+d\bar r^2+\frac{\bar r^2}{4}\Bigl[(d\psi-\cos\theta\, d\phi)^2+d\theta^2+\sin^2\theta\, d\phi^2\Bigr],
\end{eqnarray}
together with the identifications
\begin{eqnarray}
(\psi,\phi)\sim(\psi+2\pi,\phi+2\pi),\qquad
(\psi,\phi)\sim(\psi+4\pi,\phi).
\end{eqnarray}

\medskip
The horizons are located at the values $\bar r$ satisfying $\Delta=0$.
The solution is extremal when $\Delta$ has a double zero in $\bar r^2$, which occurs for
\begin{equation}
m=(|a_1|+|a_2|)^2.
\end{equation}
\medskip
Taking the extremal limit 
\begin{eqnarray}
m=(a_1-a_2)^2, \label{eq:ext}
\end{eqnarray}
and introducing the new radial coordinate $r$,
 \begin{eqnarray}
\bar r=\sqrt{r-4a_1a_2},
\end{eqnarray}
together with the new parameters
\begin{eqnarray}
a_\pm=\frac{a_1\pm a_2}{2}, \label{eq:apm}
\end{eqnarray}
the metric can be rewritten in the form
\begin{eqnarray}
ds^2&=&\frac{2 r^2+2 a_-^2 r -a_-^3 a_+ \cos \theta+a_-^4}{2 r+a_- a_+ \cos \theta+a_-^2}
\biggl[d\psi -\frac{a_-^2 (a_--a_+ \cos \theta)}{2 r^2+2 a_-^2 r-a_-^3 a_+ \cos \theta+a_-^4}\left( dt+\frac{a_-^2 a_+ \sin ^2\theta}{2 r}d\phi     \right)\notag \\
&& +\left(  -\cos\theta+\frac{a_-a_+ \sin ^2\theta}{2 r}\right)d\phi           \biggr]^2 -\frac{2 r^2}{2 r^2+2 a_-^2 r-a_-^3 a_+ \cos \theta+a_-^4}
\left(
dt+\frac{a_-^2 a_+ \sin ^2\theta}{2 r}d\phi
\right)^2\notag \\
&&+\frac{2r+a_-^2+a_- a_+ \cos \theta}{2 r^2}[dr^2+r^2(d\theta^2+\sin^2\theta d\phi^2)]. \label{eq:exMP}
\end{eqnarray}

\medskip

For this  extremal Myers-Perry solution, the metric $h_{ij}$ takes the flat form:
\begin{eqnarray}
h_{ij}dx^idx^j=dr^2+r^2(d \theta^2+\sin^2\theta d\phi^2),
\end{eqnarray}
and the corresponding coset matrix $\chi$~(\ref{eq:chi}) can be written as the form 
depending on two harmonic functions  $f$ and $g$ on the flat space:
\begin{eqnarray}
\chi=\eta e^{fA}e^{gA^2}, \label{eq:chi_MP}
\end{eqnarray}
with
\begin{eqnarray}
f=\frac{1}{r}, \quad g=\frac{a_+}{2a_-}\frac{\cos\theta }{r^2}, \label{eq:fg1}
\end{eqnarray}
where 
the matrices $\eta$, $A$ are given by
\begin{equation}
\eta=\left(
\begin{array}{ccc}
 -1 & 0 & 0 \\
 0 & 0 & -1 \\
 0 & -1 & 0 \\
\end{array}
\right),
\end{equation}
\begin{eqnarray}
A=
\begin{pmatrix}
 -a_-^2 &  a_-^3&  0 \\
  0 &  \frac{1}{2}a_-^2 &1  \\
 a_-^3&-\frac{3}{4}a_-^4&   \frac{1}{2}a_-^2
\end{pmatrix}.
 \label{eq:A}
\end{eqnarray}
where it should be emphasized that, in constructing the scalar fields $\{\lambda_{ab},V_a\}$, the pair of Killing vectors $\xi_a$ must be chosen as $\xi_0=\partial_t$ and $\xi_1=\partial_\psi$, rather than $\{\xi_0,\xi_1\}=\{\partial_t,\partial_{\phi_1}\}$ or $\{\xi_0,\xi_1\}=\{\partial_t,\partial_{\phi_2}\}$.

\medskip
In what follows, we demonstrate this explicitly.
From  (\ref{eq:fg1})--(\ref{eq:A}), the coset matrix $\chi$ takes the form
\begin{eqnarray}
\setlength{\arraycolsep}{1pt}  
\chi
=
\left(
\begin{array}{ccc}
 \displaystyle-\frac{1}{2}a_{-}^4 \left(f^2+2 g\right)+a_{-}^2 f-1 &\displaystyle \frac{1}{4} \left[a_{-}^5 \left(f^2+2 g\right)-4a_{-}^3 f\right] & \displaystyle-\frac{1}{2}a_{-}^3 \left(f^2+2 g\right) \\
\displaystyle \frac{1}{4} \left[ a_{-}^5 \left(f^2+2 g\right)-4a_{-}^3 f\right] &\displaystyle \frac{1}{8} \left[6a_{-}^4 f-a_{-}^6 \left(f^2+2 g\right)\right] &\displaystyle \frac{1}{4} \left[a_{-}^4 \left(f^2+2 g\right)-2a_{-}^2 f-4\right] \\
\displaystyle -\frac{1}{2}a_{-}^3 \left(f^2+2 g\right) &\displaystyle \frac{1}{4} \left[a_{-}^4 \left(f^2+2 g\right)-2a_{-}^2 f-4\right] & \displaystyle\frac{1}{2} \left[-a_{-}^2(f^2+2g)-2 f\right] \\
\end{array}
\right).
\end{eqnarray}
Hence, from Eq.~(\ref{eq:chidef}), we can read off the conformal factor $\tau$ and the scalar fields $(\lambda_{ab},V_a)$ as
\begin{eqnarray}
\tau&=&\frac{2}{a_{-}^2( f^2+2g)+2 f},\\
\lambda_{00}&=&\frac{\tau}{2}[a_{-}^2( f^2-2g)-2 f],\\
\lambda_{01}&=&-\frac{\tau}{2} \left[a_{-}^3 (f^2 - 2 g)\right],\\
\lambda_{11}&=&\frac{\tau}{2} \left[a_{-}^4 \left(f^2-2 g\right)+2 a_{-}^2 f+2\right] ,\\
V_0&=& -\frac{\tau}{2}[a_{-}^3 \left(f^2+2 g\right)], \label{eq:V0}\\
V_1&=&\frac{\tau}{4}[a_{-}^4 \left(f^2+2 g\right)-2 a_{-}^2 f-4]. \label{eq:V5}
\end{eqnarray}
From Eqs.~(\ref{eq:domega}), (\ref{eq:V0}) and (\ref{eq:V5}), the 1-forms ${\bm \omega}^0=\omega^0{}_i dx^i$ and ${\bm \omega}^1=\omega^1{}_i dx^i$ can be expressed as
\begin{eqnarray}
{\bm \nabla} \times {\bm \omega}^0&=&-a_{-}^3{\bm \nabla} g \label{eq:omega0a},\\
{\bm \nabla} \times {\bm \omega}^1&=& -{\bm \nabla}\left( f+a_{-}^2 g \right).\label{eq:omega5}
\end{eqnarray}
If we define $\tilde {\bm \omega}^1:={\bm \omega^1}-\frac{1}{a_{-}}{\bm \omega^0}$, then
\begin{eqnarray}
{\bm \nabla} \times \tilde {\bm \omega}^1=- {\bm \nabla}  f.  \label{eq:omega5t}
\end{eqnarray}
From Eq.~(\ref{eq:fg1}),  these can be solved as
\begin{eqnarray}
{\bm \omega}^0&=&\frac{a_{-}^2a_+}{2r}\sin^2\theta d\phi=-\frac{a_{-}^2a_+}{2r^3}[ydx-xdy],\\
 \tilde{\bm \omega}^1&=& -\cos\theta d\phi=\frac{z}{r}\frac{ydx-xdy}{x^2+y^2},
\end{eqnarray}
with $(x,y,z)=(r\sin\theta\cos\phi,r\sin\theta\sin\phi,r\cos\theta)$. Thus  the 5D metric~(\ref{eq:5D metric})  can be obtained as
\begin{eqnarray}
ds^2&=&\frac{H_-}{H_+}
\left[d\psi- \frac{a_{-}^3(f^2-2g)}{2H_-} \left(dt+\omega^0_\phi d\phi \right)
+\omega^1_\phi d\phi \right]^2\notag\\
&&-\frac{1}{H_-}\left (dt+\omega^0_\phi d\phi \right )^2+H_+\left[dr^2+r^2(d\theta^2+\sin^2\theta d\phi^2)\right],
\end{eqnarray}
where the functions $H_\pm$ are given  by
\begin{eqnarray}
H_-&=&\frac{1}{2}[a_{-}^4(f^2-2g)+2a_{-}^2f+2],\notag \\
H_+&=&\frac{1}{2}[a_{-}^2(f^2+2g)+2f].
\label{eq:hpm}
\end{eqnarray}
This coincides with the metric~(\ref{eq:exMP}) corresponding to  the extremal Myers-Perry black hole. 

\subsection{Multi-centered rotating black hole solutions}

 We now generalize the harmonic functions $f$ and $g$ in Eq.~(\ref{eq:fg1}) to multi-center configurations by taking
\begin{eqnarray}
f&=&\sum_{i=1}^N\frac{1}{|{\bm x}-{\bm x_i}|},  \label{eq:fm}\\
g&=&\sum_{i=1}^N\frac{j_i(z-z_i)}{|{\bm x}-{\bm x_i}|^3}, \label{eq:gm}
\end{eqnarray}
where $j_i$ are new parameters.
Using  Eqs.~(\ref{eq:omega0a}) and (\ref{eq:omega5t}), we can show that the one-forms ${\bm \omega}^0$ and $\tilde {\bm \omega}^1$ take the following explicit forms:
\begin{eqnarray}
{\bm \omega}^0&=&-\sum_{i=1}^Na_-^3j_i\frac{(y-y_i)dx-(x-x_i)dy}{|{\bm x}-{\bm x_i}|^3},  \label{eq:omega0}\\
 \tilde{\bm \omega}^1&=& \sum_{i=1}^N\frac{(z-z_i)}{|{\bm x}-{\bm x_i}|}\frac{(y-y_i)dx-(x-x_i)dy}{(x-x_i)^2+(y-y_i)^2}.\label{eq:tomega5}
\end{eqnarray}
Consequently, the 5D metric of the resulting multi-centered black hole solution can be written as
\begin{eqnarray}
ds^2&=&\frac{H_-}{H_+}
\left[d\psi- \frac{a_{-}^3(f^2-2g)}{2H_-} \left(dt+\bm{\omega^0} \right)
+{\bm  \omega}^1\right]^2-\frac{1}{H_-}\left (dt+\bm{\omega}^0\right )^2+H_+d{\bm x}\cdot d{\bm x},\quad 
{\bm x}=(x,y,z)\label{eq:sol}
\end{eqnarray}
together with Eqs.~(\ref{eq:hpm}), (\ref{eq:fm}), (\ref{eq:gm}), (\ref{eq:omega0}), and (\ref{eq:tomega5}).
As we shall see in the next section, this metric describes a family of multi-centered rotating black holes.

\section{Properties of the multi-rotating black hole solution}\label{sec:anaysis}

In this section, we see that this solution is regular and describes asymptotically flat, multi-rotating black holes, each possessing an extremal horizon with the topology of $S^3$.
We demonstrate that curvature singularities are confined inside the horizons and do not occur on or outside them.
Furthermore, we prove the absence of closed timelike curves (CTCs) in the exterior region as well as on the horizons.

\subsection{Near-horizon geometry}

The metric apparently diverges at the point sources ${\bm x}={\bm x}_i$ in the harmonic functions $f$ and $g$  but we show that they correspond to smooth Killing horizons, provided  for all $i$, the parameters $j_i$ satisfy
\begin{eqnarray}
|j_i|<\frac{1}{2}.
\end{eqnarray}
Using the new radial coordinate $r:=|{\bm x}-{\bm x}_i|$ and the spherical coordinates $(x,y,z)=(r\sin\theta \cos\phi,r\sin\theta\sin\phi,r\cos\theta)$  so that the $i$-th point source becomes an origin, we examine the behavior of the metric near $r=0$.
First, we introduce new coordinates $(t',r',\psi',\phi')$ defined by
\begin{eqnarray}
d\psi=d\psi'+(A_0-B_0)dt,\quad 
d\phi=d\phi'+(A_0+B_0)dt,\quad 
dt=\frac{E}{\varepsilon}dt',\quad
dr=\varepsilon D dr',
\end{eqnarray}
with the constants 
\begin{eqnarray}
A_0=-B_0=\frac{1}{2a_-},\quad D=4a_-^2,\quad E=\frac{\sqrt{a_-^2-a_+^2}}{2}
\end{eqnarray}
Next, we introduce the coordinate $(v,\psi'',\phi'')$ by
\begin{eqnarray}
dt'=dv+\left(\frac{A_2}{r^{\prime 2}}+\frac{A_1}{r'}\right)dr',\quad 
d\psi'=d\psi''+\frac{B_1}{r'}dr', \quad 
d\phi'=d\phi''+\frac{C_1}{r'}dr'
\end{eqnarray}
with
\begin{eqnarray}
A_1&=&\pm\frac{a_-(1+a_-^2 f_s)\varepsilon}{E\sqrt{1-4 j_i^2}}=\pm\frac{2a_-(1+a_-^2 f_s)\varepsilon}{\sqrt{a_-^2-a_+^2}\sqrt{1-4 j_i^2}},\\
A_2&=&\pm\frac{a_-^3 \sqrt{1-4 j_i^2}}{2 D E}=\pm\frac{a_-\sqrt{1-4j_i^2}}{4\sqrt{a_-^2-a_+^2}},\\
B_1&=&\mp\frac{1}{\sqrt{1-4 j_i^2}}\\
C_1&=&\pm\frac{2j_i}{\sqrt{1-4 j_i^2}}.
\end{eqnarray}
then, the apparent divergence can be eliminated. 
Finally, taking the near-horizon limit $\varepsilon\to 0$, we obtain the metric
\begin{eqnarray}
ds^2&\simeq& \frac{2D^2E^2}{a_-^4f_+}r^{\prime 2}dv^2\mp\frac{2DEf_+}{a_-\sqrt{1-4j_i^2}}dvdr'+\frac{4 DE }{a_-f_+}r'dv[d\psi''- (\cos\theta -j_i \sin^2\theta) d\phi''   ]\notag \\
&&+\frac{a_-^2f_-}{f_+}\left[d\psi''+\frac{2j_i-\cos\theta}{f_-}d\phi''\right]^2+\frac{a_-^2(1-4j_i^2)}{2f_-}\sin^2\theta d{\phi''}^2+\frac{a_-^2f_+}{2}d\theta^2,
\end{eqnarray}
where $f_\pm:=1\pm 2j_i\cos\theta$. 
The horizon area is
\begin{eqnarray}
A=8\pi a_-^3\sqrt{1-4j_i^2}.
\end{eqnarray}
The nonvanishing horizon area together with the absence of CTCs near the horizon requires the conditions:
\begin{eqnarray}
a_->0,\quad |j_i|<\frac{1}{2}, \quad i=1,\ldots,N.
\end{eqnarray}
This is the same near-horizon geometry as that of an extremal Myers--Perry black hole whose horizon cross section is $S^3$, provided the angular coordinates satisfy the periodicities $\Delta \psi = 4\pi$ and $\Delta \phi = 2\pi$, equivalently $\Delta \psi'' = 4\pi$ and $\Delta \phi'' = 2\pi$.

\subsection{Asymptotic structure}\label{sec:asymp}
In terms of the standard spherical coordinates $(x,y,z)=(r\sin\theta \cos\phi,r\sin\theta\sin\phi,r\cos\theta)$, 
the functions, $f,g,H_\pm$ and the $1$-forms ${\bm \omega}^0$, $\tilde {\bm \omega}^1$  at $r\to \infty$
  behave asymptotically as\begin{eqnarray}
f&\simeq& \frac{N}{r}+{\cal O}(r^{-2}),\\
g&\simeq& \frac{\sum_ij_i\cos\theta}{r^2}+{\cal O}(r^{-3}),\\
H_+&\simeq& \frac{N}{r}+{\cal O}(r^{-2}),\\
H_-&\simeq& 1+\frac{Na_-^2}{r}+{\cal O}(r^{-2}),\\
\end{eqnarray}
and 
\begin{eqnarray}
{\bm \omega}^0&\simeq& \left(\frac{a_-^3 \sum_i j_i}{r}\sin^2\theta+{\cal O}(r^{-2})\right)d\phi,\\
\tilde {\bm \omega}^1&\simeq&\left(-N\cos\theta+{\cal O}(r^{-1})\right)d\phi.
\end{eqnarray}

In terms of  the new radial coordinate $\tilde r:=2\sqrt{Nr}$, the metric  at $\tilde r\to\infty$ behaves as 
\begin{eqnarray}
ds^2&\simeq&-dt^2+d\tilde r^2 +\frac{\tilde r^2}{4}\left[\left(\frac{d\psi}{N} -\cos\theta d\phi \right)^2+d\theta^2+\sin^2\theta d\phi^2\right],
\end{eqnarray}
which   is locally isometric to 5D Minkowski spacetime, with the spatial infinity $S^3$ replaced with the lens space $L(N;1)=S^3/{\mathbb Z}_N$.

\subsection{Regularity}

If curvature singularities exist outside the horizons, they appear at points where the metric or its inverse diverges, which happens only on the surfaces $H_+(x,y,z)=0$ or $H_-(x,y,z)=0$.
We can show that such singularities do not exist on and outside the event horizons at ${\bm x}={\bm x}_i$ provided that $|j_i|<1/2$.
To demonstrate this, it is sufficient to verify that $H_\pm>0$ on and outside the horizons. 
First, we note that under the assumptions, we have $f>0$, while $g$ can take both signs. 

\medskip
We  normalized all quantities $Q$  by the mass $m$, and we denote the corresponding dimensionless quantities by $\hat Q$. 
For example, 
\begin{eqnarray}
\hat{\bm x}=\frac{{\bm x}}{m}, \quad\hat{\bm x_i}=\frac{{\bm x}_i}{m},\quad \hat f=m f=\sum_i\frac{1}{|\hat{\bm x}-\hat{\bm x}_i|},\quad \hat g=m^2 g=\sum_i\frac{j_i(\hat z-\hat z_i)}{|\hat{\bm x}-\hat{\bm x}_i|^2}
\end{eqnarray}
and, it follows from eqs.(\ref{eq:ext}) and (\ref{eq:apm}) that
\begin{equation}
\hat a_-=\frac{a_-}{\sqrt{m}}=\frac{1}{2}, \quad \hat j_i=j_i.
\end{equation}

Using this, we can  show the positivity of the function $\hat H_-=H_-$ as
\begin{eqnarray}
2\hat H_-   &=& \hat a_-^4(\hat f^2-2\hat g)+2\hat a_-^2\hat f+2\notag\\
           &>& \hat a_-^4[(\hat f^2-2\hat g)+2\hat f+2]\notag\\
            &=& \hat a_-^4[(1+\hat f)^2-2\hat g+1]\notag\\
             &\ge & \hat a_-^4[(1+\hat f)^2-2|\hat g|+1]\notag\\
             &>&0 \label{eq:Hm>0}
\end{eqnarray}
where we have used the inequality
\begin{eqnarray}
(1+\hat f)^2-2|\hat g|&=&\left(1+\sum_i \hat f_i\right)^2-2|\sum_i \hat g_i|\notag\\
                  &>&1+\sum_i \hat f_i^2-2\sum_i |\hat g_i|\notag\\
                  &\ge &1+\sum_i \frac{1-2|j_i|}{|\hat {\bm x}-\hat {\bm x}_i|^2}\notag\\
                  &>&1 \label{eq:ineq1}
\end{eqnarray}
with
\begin{eqnarray}
\hat f_i:=\frac{1}{|\hat{\bm x}-\hat{\bm x_i}|},\quad \hat g_i:=\frac{j_i (\hat z-\hat z_i)}{|\hat{\bm  x}-\hat {\bm x_i}|^3}.
\end{eqnarray}
Furthermore, we can  show the positivity of the function $\hat H_+:=mH_+$ as follows:
\begin{eqnarray}
2\hat H_+   &=& \hat a_-^2(\hat f^2+2\hat g)+2\hat f\notag\\
           &\ge &\hat a_-^2[(\hat f^2+2\hat g)+2\hat f]\notag\\
            &=& \hat a_-^2[(1+\hat f)^2+2\hat g-1]\notag\\
             &\ge & \hat a_-^2[(1+\hat f)^2-2|\hat g|-1]\notag\\
             &>&0, \label{eq:Hp>0}
\end{eqnarray}
where we have used the inequality~(\ref{eq:ineq1}).

\subsection{Absence of CTCs}

Here, we show the nonexistence of CTCs everywhere on and outside the horizons, provided that the inequality $|j_i| < 1/2$ holds for each $i = 1, \ldots, N$.
The condition for the absence of CTCs is equivalent to requiring that the $3$D matrix
$g_{IJ}$ $(I, J =\psi, x, y)$ is positive definite everywhere on and outside the horizons. 
From the Sylvester's criterion, the 3D matrix $(g_{IJ})$ is  positive-definite if and only if all the leading principal minors of the matrix $(g_{IJ})$ are positive as follows: 
\begin{eqnarray}
g_{\psi\psi}&=&\frac{H_-}{H_+}>0, \label{eq:ineq1b}\\
{\rm det} 
\begin{pmatrix}
g_{\psi\psi} & g_{\psi x} \\
g_{\psi x} & g_{xx} \\
\end{pmatrix}
&=&\frac{H_+H_--(\omega^0_x)^2}{H_+}>0,  \label{eq:ineq2b}\\
{\rm det} 
\begin{pmatrix}
g_{\psi\psi} & g_{\psi x}&g_{\psi y}\\
g_{\psi x} & g_{xx} & g_{xy}\\
g_{\psi y} & g_{xy} & g_{yy}\\
\end{pmatrix}
&=&H_+H_--(\omega^0_x)^2-(\omega^0_y)^2>0. \label{eq:ineq3b}
\end{eqnarray}
From the inequalities (\ref{eq:Hm>0}) and (\ref{eq:Hp>0}), the condition~(\ref{eq:ineq1b}) is satisfied.
It is straightforward to verify that if the condition~(\ref{eq:ineq3b}) is satisfied, then the condition~(\ref{eq:ineq2b}) is also automatically satisfied. Therefore, it is sufficient to consider only the inequality~(\ref{eq:ineq3b}).

\medskip
Using the inequalities~(\ref{eq:Hm>0}) and (\ref{eq:Hp>0}) 
\begin{eqnarray}
2\hat H_+\ge \hat a_-^2[(1+\hat f)^2+2\hat g-1]>0,\\
2\hat H_-\ge \hat a_-^4[(1+\hat f)^2-2\hat g+1]>0,
\end{eqnarray}
we have
\begin{eqnarray}
4[\hat H_+\hat H_--(\hat \omega^0_x)^2 - (\hat \omega^0_y)^2]&\ge &  \hat a_-^6\left[ \{(1+\hat f)^2+2\hat g-1\}\{(1+\hat f)^2-2\hat g+1\}-4\hat a_-^{-6}(\hat \omega^0_x)^2 - 4\hat a_-^{-6}(\hat \omega^0_y)^2 \right] \notag\\
&=&\hat a_-^6\left[ (1+\hat f)^4-(2\hat g-1)^2-4\hat a_-^{-6}(\hat \omega^0_x)^2 - 4\hat a_-^{-6}(\hat\omega^0_y)^2 \right] \notag\\
&=&\hat a_-^6\left[ 
\left(1+\sum_i \hat f_i\right)^4-\left(2\sum_i\hat g_i-1\right)^2-4\left(\sum_i \check \omega^0_{x,i}\right)^2 - 4\left(\sum_i\check \omega^0_{y,i}\right)^2
\right] \notag \\
&\ge&\hat a_-^6\biggl[ \sum_i \left(\hat f_i^4-4\hat g_i^2-4(\check \omega^0_{x,i})^2-4(\check \omega^0_{y,i})^2\right)+\sum_i\left(2\hat f_i^2-4|\hat g_i|\right)
\notag \\
&&+\sum_{i\not=j}\left(
3\hat f_i^2 \hat f_j^2-4\hat g_i \hat g_j-4\check \omega^0_{x,i}\check \omega^0_{x,j}-4\check \omega^0_{y,i}\check \omega^0_{y,j}
\right) \biggr]\notag\\
\end{eqnarray}
with
\begin{eqnarray}
\check \omega^0_{x,i}:=\hat a_-^{-3}\hat \omega^0_{x,i}=\frac{j_i(\hat y-\hat y_i)}{|\hat{\bm x}-\hat {\bm x_i}|^3},\quad 
\check \omega^0_{y,i}:=\hat a_-^{-3}\hat \omega^0_{y,i}=\frac{j_i(\hat x-\hat x_i)}{|\hat{\bm x}-\hat{\bm x_i}|^3},
\end{eqnarray}
where the first, second and third summations can be shown to be positive under the conditions $|j_i|<1/2\ (i=1,\ldots,N)$. Indeed,
\begin{eqnarray}
\hat f_i^4-4\hat g_i^2-4(\check \omega^0_{x,i})^2-4(\check \omega^0_{y,i})^2
&=&\frac{1}{|\hat{\bm x}-\hat{\bm x}_i|^4}-\frac{4j_i^2(\hat z-\hat z_i)^2}{|\hat{\bm x}-\hat{\bm x}_i|^6}
-\frac{4j_i^2(\hat y-\hat y_i)^2}{|\hat{\bm x}-\hat{\bm x}_i|^6}
-\frac{4j_i^2(\hat x-\hat x_i)^2}{|\hat{\bm x}-\hat{\bm x}_i|^6}\notag\\
&=&\frac{1-4j_i^2}{|\hat{\bm x}-\hat{\bm x}_i|^4}\notag\\
&>&0,
\end{eqnarray}
\begin{eqnarray}
2\hat f_i^2-4|\hat g_i|&=&2\frac{1}{|\hat{\bm x}-\hat{\bm x_i}|^2}-4\frac{|j_i||\hat z-\hat z_i|}{|\hat{\bm x}-\hat{\bm x_i}|^3}\\
&\ge &2\frac{1}{|\hat{\bm x}-\hat{\bm x_i}|^2}-4\frac{|j_i|}{|\hat{\bm x}-\hat{\bm x_i}|^2}\notag \\
&=&2\frac{1-2|j_i|}{|\hat{\bm x}-\hat{\bm x_i}|^2}\notag\\
&>&0,
\end{eqnarray}
and
\begin{eqnarray}
&&3\hat f_i^2\hat f_j^2-4\hat g_i\hat g_j-4\check \omega^0_{x,i}\check \omega^0_{x,j}-4\check \omega^0_{y,i}\check \omega^0_{y,j}\notag\\
&&=\frac{3}{|\hat{\bm x}-\hat{\bm x}_i|^2|\hat{\bm x}-\hat{\bm x}_j|^2}-\frac{4j_ij_j (\hat z-\hat z_i)(\hat z-\hat z_j)}{|\hat{\bm x}-\hat{\bm x}_i|^3|\hat{\bm x}-\hat{\bm x}_j|^3}
-\frac{4j_ij_j[(\hat x-\hat x_i)(\hat x-\hat x_j)+(\hat y-\hat y_i)(\hat y-\hat y_j)]}{|\hat{\bm x}-\hat{\bm x_i}|^3|\hat{\bm x}-\hat{\bm x}_j|^3}\notag\\
&&=\frac{3}{|\hat{\bm x}-\hat{\bm x}_i|^2|\hat{\bm x}-\hat{\bm x}_j|^2}-\frac{4j_ij_j(\hat{\bm x}-\hat{\bm x}_i)\cdot(\hat{\bm x}-\hat{\bm x}_j)}{|\hat {\bm x}-\hat{\bm x}_i|^3|\hat{\bm x}-\hat{\bm x}_j|^3}\notag\\
&&\ge\frac{3}{|\hat{\bm x}-\hat{\bm x}_i|^2|\hat{\bm x}-\hat{\bm x}_j|^2}-\frac{12|j_ij_j|}{|\hat {\bm x}-\hat{\bm x}_i|^2|\hat{\bm x}-\hat{\bm x}_j|^2}\notag\\
&&=\frac{3(1-4| j_i||j_j|)}{|\hat{\bm x}-\hat{\bm x}_i|^2|\hat{\bm x}-\hat{\bm x}_j|^2}\notag\\
&&>0.
\end{eqnarray}
The positivity of these equations implies that ${\rm det\ }g^{(4)}|_{(x,y)}>0$.
Therefore, no CTCs exist on or outside the horizons.

\section{Myers-Perry black hole binary}\label{sec:binary}
The symmetry of the multi-black hole solution is enhanced when all black holes are  placed on the $z$-axis, since the geometry then admits  $U(1)\times U(1)$ isometry due to an additional $U(1)$ isometry generated by the rotation about the  axis. 
We now give the explicit form of the solution in the case of two black holes.
Without loss of generality, we take their positions to be ${\bm x}_1=(0,0,-a)$
and ${\bm x}_2=(0,0,a)$. Then, introducing cylindrical coordinates $(\rho,\phi,z)$ defined by $(x,y,z)=(\rho\cos\phi,\rho \sin \phi,z)$, we have 
\begin{eqnarray}
ds^2&=&\frac{H_-}{H_+}
\left[d\psi- \frac{a_{-}^3(f^2-2g)}{2H_-} \left(dt+\omega^0{}_\phi d\phi\right)
+\omega^1{}_\phi d\phi\right]^2-\frac{1}{H_-}\left (dt+\omega^0{}_\phi d\phi\right )^2\notag\\
&&+H_+(d\rho^2+dz^2+\rho^2d\phi^2),
\label{eq:sol_2MP}
\end{eqnarray}
with (\ref{eq:hpm}) and 
\begin{eqnarray}
f&=&\frac{1}{\sqrt{\rho^2+(z+a)^2}}+\frac{1}{\sqrt{\rho^2+(z-a)^2}},\label{eq:f_EH}\\
g&=&\frac{j_1(z+a)}{\sqrt{\rho^2+(z+a)^2}^3}+\frac{j_2(z-a)}{\sqrt{\rho^2+(z-a)^2}^3},\\
\omega^0{}_\phi&=&\frac{a_-^3j_1\rho^2}{\sqrt{\rho^2+(z+a)^2}^3}+\frac{a_-^3j_2\rho^2}{\sqrt{\rho^2+(z-a)^2}^3},\\
\tilde  \omega^1{}_\phi&=&-\frac{z+a}{\sqrt{\rho^2+(z+a)^2}}-\frac{z-a}{\sqrt{\rho^2+(z-a)^2}},\label{eq:tomega1_EH}
\end{eqnarray}
where  $\partial_\phi$ is the rotational Killing vector around the $z$-axis, together with the rotational Killing vector $\partial_\psi$.
As shown in Sec.~\ref{sec:asymp},
the asymptotic structure of the spacetime is locally 5D  Minkowski spacetime, and the spatial infinity has the topology of the lens space $L(2;1)=S^3/{\mathbb Z}_2$. 
 This type of asymptotic structure for the timeslice $t=$constant is referred to as asymptotically locally Euclidean (ALE), as in the case of the Eguchi-Hanson space~\cite{Eguchi:1978xp,Eguchi:1978gw}. 
  Multi-black hole solutions~\cite{Ishihara:2006pb} and black ring solutions~\cite{Tomizawa:2007he,Tomizawa:2008tj} on the Eguchi-Hanson space have been constructed as BPS solutions  in 5D minimal supergravity (Einstein-Maxwell-Chern-Simons theory).  
 In fact, in the limit where the rotation parameters $a_-$ and $j_i$ $(i=1,2)$
vanish, the functions $\omega^0{}_\phi$ and $g$ also vanish, and the metric
reduces to
\begin{eqnarray}
ds^2 &=& -dt^2 + ds_4^2, \\
&ds_4^2& = f^{-1}(d\psi+\tilde{\omega}^1{}_\phi\, d\phi)^2
          + f(d\rho^2+dz^2+\rho^2 d\phi^2),\notag
\end{eqnarray}
where $f$ and $\tilde{\omega}^1{}_\phi$ are given by
Eqs.~(\ref{eq:f_EH}) and (\ref{eq:tomega1_EH}), respectively.
The spatial metric $ds_4^2$ coincides with the Eguchi--Hanson metric
written in the Gibbons--Hawking form~\cite{Eguchi:1978xp,Prasad:1979kg}.
Except for the horizons $z=\pm a$, the $z$-axis corresponds to the rotational axis, where the spacelike Killing vectors with closed integral curves vanish.

\medskip
Below, to show the absence of conical singularities on the $z$-axis, we
analyze the rod structure of the spacetime.
Let $\eta$ be an angular coordinate with period $\Delta\eta$, associated with a
rotational Killing vector $\partial_\eta$.
Then regularity on the axis (i.e.\ the absence of conical singularities)
requires the condition~\cite{Harmark:2004rm}
\begin{eqnarray}\label{eq:conical-free}
\Delta \eta=2\pi \lim_{\rho\to 0}\sqrt{ \frac{\rho^2 g_{\rho\rho} }{g(\partial_\eta,\partial_\eta)}}.
\end{eqnarray}
To investigate the rod structure of this solution, it is more convenient to use the angular coordinate $\Psi=\psi/2$.
Then, the metric  at $\tilde r\to\infty$ behaves as 
\begin{eqnarray}
ds^2&\simeq&-dt^2+d\tilde r^2 +\frac{\tilde r^2}{4}\left[\left(d\Psi -\cos\theta d\phi \right)^2+d\theta^2+\sin^2\theta d\phi^2\right].
\end{eqnarray}
This $z$-axis can be divided into three rod:
\begin{itemize}
\item[(i)] the $\phi_-$-rotational axis: $\Sigma_-=\{(\rho,z)| \rho=0,z<-a \}$,  with direction $\ell_-=\partial_{\phi_-}:=-2\partial_\psi+\partial_\phi=-\partial_{\Psi}+\partial_\phi$.

\item[(ii)]  the $\phi$-rotational axis:  $\Sigma_\phi=\{(\rho,z)| \rho=0,-a<z<a \}$,  with direction $\ell=\partial_\phi$

\item[(iii)]  the $\phi_+$-rotational axis: $\Sigma_+=\{(\rho,z)| \rho=0,z>a \}$, with direction $\ell_+=\partial_{\phi_+}:=2\partial_\psi+\partial_\phi=\partial_{\Psi}+\partial_\phi$.
\end{itemize}
This rod structure is the same as that of the Eguchi--Hanson space discussed in~\cite{Chen:2010zu}, except that our solution is 5D Lorentzian and contains horizons.
In fact, the coordinates $(\phi,\Psi,\phi_-,\phi_+,z)$ used in this paper correspond to $(\psi,\phi,\tilde{\psi},\tilde{\phi},-z)$ in~\cite{Chen:2010zu}.

\medskip
To ensure regularity, we assume that the orbits generated by $\{\ell_+,\ell\}$ are independently identified with period $2\pi$, which implies
\begin{eqnarray}
(\phi_+,\phi)\sim(\phi_+,\phi+2\pi),\qquad
(\phi_+,\phi)\sim(\phi_++2\pi,\phi).
\end{eqnarray}
Since the pair $\{\ell_+-\ell,\ell\}=\{\partial_{\Psi},\partial_\phi\}$ is related to $\{\ell_+,\ell\}$ by a $GL(2,\mathbb{Z})$ transformation, it can equally well be taken as a pair of independent $2\pi$-periodic generators. Therefore,
\begin{eqnarray}
(\Psi,\phi)\sim(\Psi,\phi+2\pi),\qquad
(\Psi,\phi)\sim(\Psi+2\pi,\phi).
\end{eqnarray}
Furthermore, since the pair $\{\ell_-,\ell\}=\{\partial_{\phi_-},\partial_\phi\}$ is related to $\{\ell,\ell_+\}$ by a $GL(2,\mathbb{Z})$ transformation, the pair $\{\ell_-,\ell\}$ may also be chosen as a set of independent generators with period $2\pi$. This leads to the identification
\begin{eqnarray}
(\phi_-,\phi)\sim(\phi_-,\phi+2\pi),\qquad
(\phi_-,\phi)\sim(\phi_-+2\pi,\phi).
\end{eqnarray}
Indeed, we can show that, on the rods $\Sigma_-$, $\Sigma_\phi$, and $\Sigma_+$, the metric~(\ref{eq:sol_2MP}) satisfies the following conditions in accordance with the above identifications:
\begin{eqnarray}
\left(\frac{\Delta \phi_-}{2\pi}\right)^2
=\lim_{\rho\to 0}\frac{\rho^2\, g_{\rho\rho}}{g(\partial_{\phi_-},\partial_{\phi_-})}
=1,
\end{eqnarray}
\begin{eqnarray}
\left(\frac{\Delta \phi}{2\pi}\right)^2
=\lim_{\rho\to 0}\frac{\rho^2\, g_{\rho\rho}}{g(\partial_{\phi},\partial_{\phi})}
=1,
\end{eqnarray}
\begin{eqnarray}
\left(\frac{\Delta \phi_+}{2\pi}\right)^2
=\lim_{\rho\to 0}\frac{\rho^2\, g_{\rho\rho}}{g(\partial_{\phi_+},\partial_{\phi_+})}
=1,
\end{eqnarray}
which implies from Eq.~(\ref{eq:conical-free}) that there are no conical singularities on any of these rods.

\medskip

In particular, on $\Sigma_\phi$, since one Killing vector $\partial_\phi$ vanishes but the other Killing  vector $\partial_\psi$ does not vanish,  $\Sigma_\phi$ is topologically cylinder. 
This structure is referred to as ``bubble".  
In the vacuum case, two uncharged black holes do not seem to be in equilibrium, but they can be balanced by the presence of a bubble region between the two horizons.

\section{Summary and Discussion}\label{sec:summary}

In this paper, we have constructed and analyzed a new class of multi-centered rotating black hole solutions in 5D vacuum Einstein gravity. 
 In the presence of two commuting Killing vector fields, the 5D Einstein theory can be reduced to an $SL(3,\mathbb{R})$ non-linear sigma model coupled with 3D gravity, where the scalar sector consists of five fields—three inner products of the Killing vectors and two twist potentials.
We have shown that the five scalar fields for the extremal Myers--Perry black hole can be written in terms of two harmonic functions on flat 3D space, and we then generalize these to multi-centered harmonic functions to obtain an explicit family of multiple extremal Myers--Perry black holes located at arbitrary points in $\mathbb{E}^3$. 
We have proved that each center corresponds to a smooth $S^3$ Killing horizon provided the rotation parameters satisfy $|j_i|<1/2$, and we have derived the near-horizon geometry and  the horizon area.
We have further established the regularity of the entire spacetime: all curvature singularities lie inside the horizons, and no singularities are present in the domain of outer communication.
 We have rigorously demonstrated the absence of closed timelike curves. 
The solutions are asymptotically flat up to $\mathbb{Z}_N$ quotient, leading to lens-space asymptotics, and we have also analyzed in detail a binary rotating black-hole configuration, clarifying its rod structure and the emergence of a bubble region. 
These results provide the first explicit construction of multiple rotating 5D black holes without supersymmetry, extending classical multi-black-hole configurations to higher
dimensions and revealing geometric features unique to 5D gravity.

\medskip
Finally, we would like to discuss a further generalization of our solution.  
One may replace the harmonic function $f$ in Eq.~(\ref{eq:fm}) with  
\[
    \sum_i \frac{N_i}{|{\bm x}-{\bm x}_i|},
\]
where each $N_i$ is an integer greater than one.  
In this case, the topology of the horizon at ${\bm x}={\bm x}_i$ becomes the lens space $L(N_i;1)$, whereas the spatial infinity has the topology of $L(\sum_i N_i;1)$.  
If, however, we take $N_i$ to be a negative integer $(N_i = -1,-2,\ldots)$, curvature singularities or CTCs may occur outside the horizon.  
In our present solution, each black hole rotates in the $(x,y)$-plane as well as in the $\psi$-direction.  
Allowing rotations in the $(y,z)$- and $(z,x)$-planes would require replacing the harmonic function $g$ in Eq.~(\ref{eq:gm}) with a more general form.  
We leave these interesting extensions for future work.

\acknowledgments
We are deeply grateful to Roberto Emparan for invaluable comments and discussions. 
ST was supported by JSPS KAKENHI Grant Number 21K03560.
RS was supported by JSPS KAKENHI Grant Number JP24K07028.




\end{document}